\documentclass[twocolumn]{el-author}

\date{6st April 2019}

\begin{document}

\title{A technique to enable frequency dependent power savings in a level crossing analog-to-digital converter}

\author{L. M. Santana, D. L. de Oliveira, L. A. Faria}

\abstract{The level crossing analog-to-digital converters are meant for the effective conversion of sparse signals by construction. In these converters, the bandwidth-power trade-off requires a re-design of the comparators which takes a lot of time and effort to reach the application optimum point. Inspired by synchronous converters that have a dynamic power component that can be traded with bandwidth with the change of a clock frequency, a technique to allow such trade-off in the level crossing converter was developed. The resulting level crossing ADC has an input signal dependent dynamic power which can reach up to 42\% OFF time during the conversion of sine waves, achieving 45.5\% power reduction in the simulated design with TSMC 180nm PDK.}

\maketitle

\section{Introduction}
The mathematical model of the level crossing conversion done by Allier \textit{et al} in \cite{1}. For this type of converter, there is no Nyquist frequency, but there are a frequency-amplitude curve which set the constraint to which the converter can still track the input signal. For a sine wave input, this relation is shown in the Inequality (\ref{eq_freq_amp}), where A is the amplitude, $f_{in}$ the frequency, $\Delta$ is the quantization level, $t_{loop}$ is the level crossing update loop time. 
\begin{align}
\label{eq_freq_amp}
    A \times f_{in} \leq \frac{\Delta}{2\pi \times t_{loop}}
\end{align}
The values of $\Delta$ and $t_{loop}$ are set during the design phase. Thus, while aiming to be efficient in the same low frequency spectrum of conversion as the SAR type ADCs, the level crossing ADC does not have the same post fabrication trade-off between bandwidth and power consumption via clock frequency control. In other words, the speed of the level crossing ADC is always over-designed for its application to the point that the total signal dependent power consumption may be inefficient. To change this scenario, a technique to power down the comparator during the time they are not being used is proposed and simulated. Being the comparators the only source of static power consumption, this technique enables power savings controlled by a clock signal which does not changes the asynchronous behavior of the architecture but does move the frequency-amplitude curve of its place.

\section{The architecture}
\begin{figure}[h!]
    \centering
    \includegraphics[width=60mm]{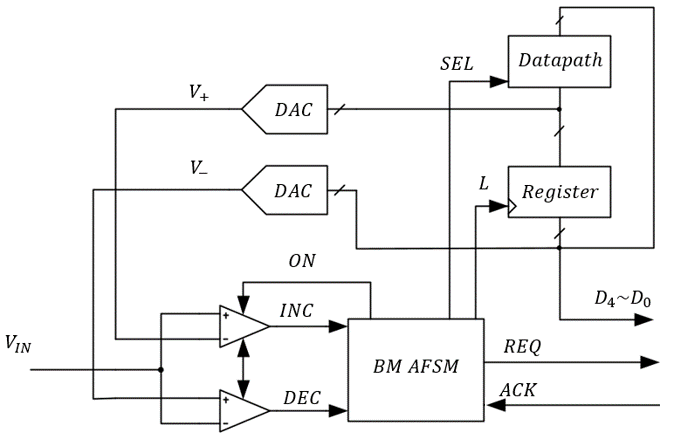}
    \caption{Modified architecture for the level crossing ADC with ON signal feeding the comparators and REQ/ACK signals closing the update loop.}
    \label{lc_adc}
\end{figure}

\begin{figure}[h!]
    \centering
    \includegraphics[width=60mm]{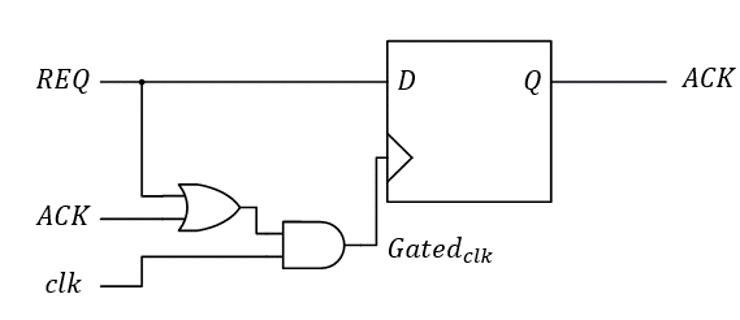}
    \caption{The circuit for controlling the off time of the comparators based on a gated clock signal.}
    \label{req_ack_circuit}
\end{figure}

\begin{figure}[h!]
    \centering
    \includegraphics[width=70mm]{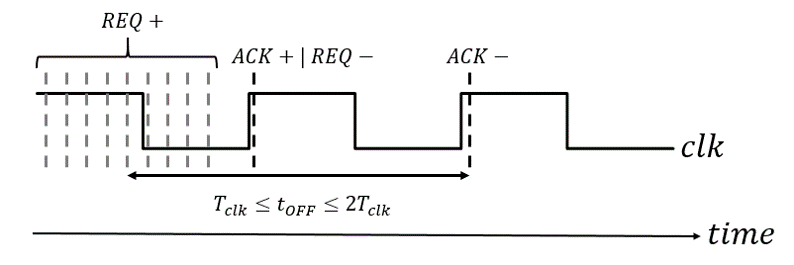}
    \caption{Timing diagram for the off time estimation based on the behaviour of the ADC and the control circuit.}
    \label{power_off_timing}
\end{figure}

The modified ADC is shown in Figure \ref{lc_adc}. The ADC level crossing loop is controlled by a Burst Mode Asynchronous Finite State Machine (BM AFSM) \cite{2}. The ADC update is event-driven by the crossing of the input signal from the floating analog window generated by the two CDACs ($V_+ - V_- = \Delta$). This state machine then updates the value of the Register which feeds the DACs with the new position for the floating window accompanying the input signal. While processing a level crossing, the comparators are powered down via an ON signal. In this architecture, this loop time can be controlled by the circuit in Figure \ref{req_ack_circuit}, which is clock gated to prevent power consumption during self-transitions \cite{4}. The Figure \ref{power_off_timing} shows a timing diagram where a REQ+ represents the level crossing event which occurs asynchronously. Then, the clock signal is used to propagate the REQ signal to the ACK signal, completing the loop of the BM AFSM. As shown, the time of this propagation is determined by the next two adjacent rising edges of the clock signal, thus it is possible to increase the loop time by choosing an appropriate slower clock. This increase in loop time will later enable the architecture to trade bandwidth for power savings.

The comparators used are continuous time comparators \cite{3}. To prevent glitches during the power up of the comparator (which would be disastrous for the BM AFSM correct behavior), these parts are modified to include two reset switches that act to discharge internal nodes of the comparators. These nodes are related to the up-transition signal path of the comparator output, thus the discharge adds to the inertia of the output and prevents the propagation of the ON signal to the output. Circuit level simulation with the TSMC 180nm PDK showed correct behavior through all corners. The average ON power of the ADC was found to be $2.6\mu W$, dominated by the static power of the comparators and the average OFF power was found to be $0.2 \mu W$, consisting of the leakage current of the transistors. 

\begin{figure}[h!]
    \centering
    \includegraphics[width=80mm]{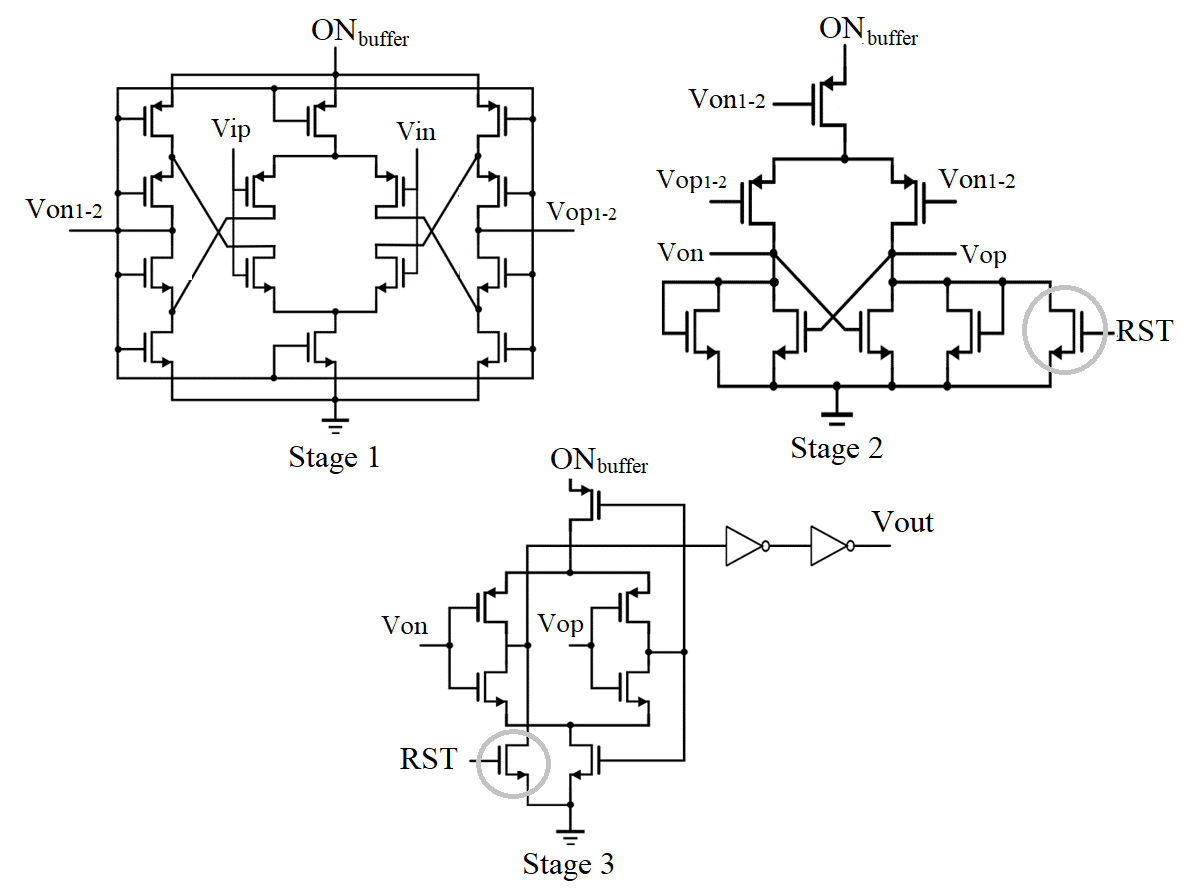}
    \caption{Three stage continuous time comparator with reset MOS to prevent glitches during power up.}
    \label{comparator}
\end{figure}

\section{Power reduction model}
Based on the timing diagram, one can derive the power reduction model for the architecture based on the assumption that the moment the level crossing occurs is random and uniformly distributed along one clock period. The Equation \ref{bandwidth} presents the boundary for the conversion of a sine wave input signal given a clock speed. The Equations \ref{t_hat} and \ref{power_red} show the estimation of the power reduction given the previous assumptions. Where A is the amplitude of a sine wave input, $f_{max}$ is the input frequency (maximum for the given amplitude), $\Delta$ is the quantization level, $T_{clk}$ is the clock period, $\hat{t}_{OFF}$ is the mean value of the off time after a conversion, $N_{cross}$ is the number of level crossing during a measurement time called $t$. As said before, the $P_{ADC}^{ON}$ is mainly dominated by the static power of the comparators and the $P_{ADC}^{OFF}$ is mainly leakage current.

\begin{equation}
    \label{bandwidth}
    f_{max} \times A  = \frac{\Delta}{2\pi \times 2T_{clk}}
\end{equation}

\begin{equation}
    \label{t_hat}
    \hat{t}_{OFF} = \frac{3 \times T_{clk}}{2}
\end{equation}

\begin{align}
    \label{power_red}
    P_{ADC}^{average} = &P_{ADC}^{ON} \times (1 - \frac{N_{cross}}{t}\times \hat{t}_{OFF}) +\nonumber\\     &P_{ADC}^{OFF} \times (\frac{N_{cross}}{t}\times \hat{t}_{OFF})
\end{align}

\begin{figure}[h!]
    \centering
    \includegraphics[width=85mm]{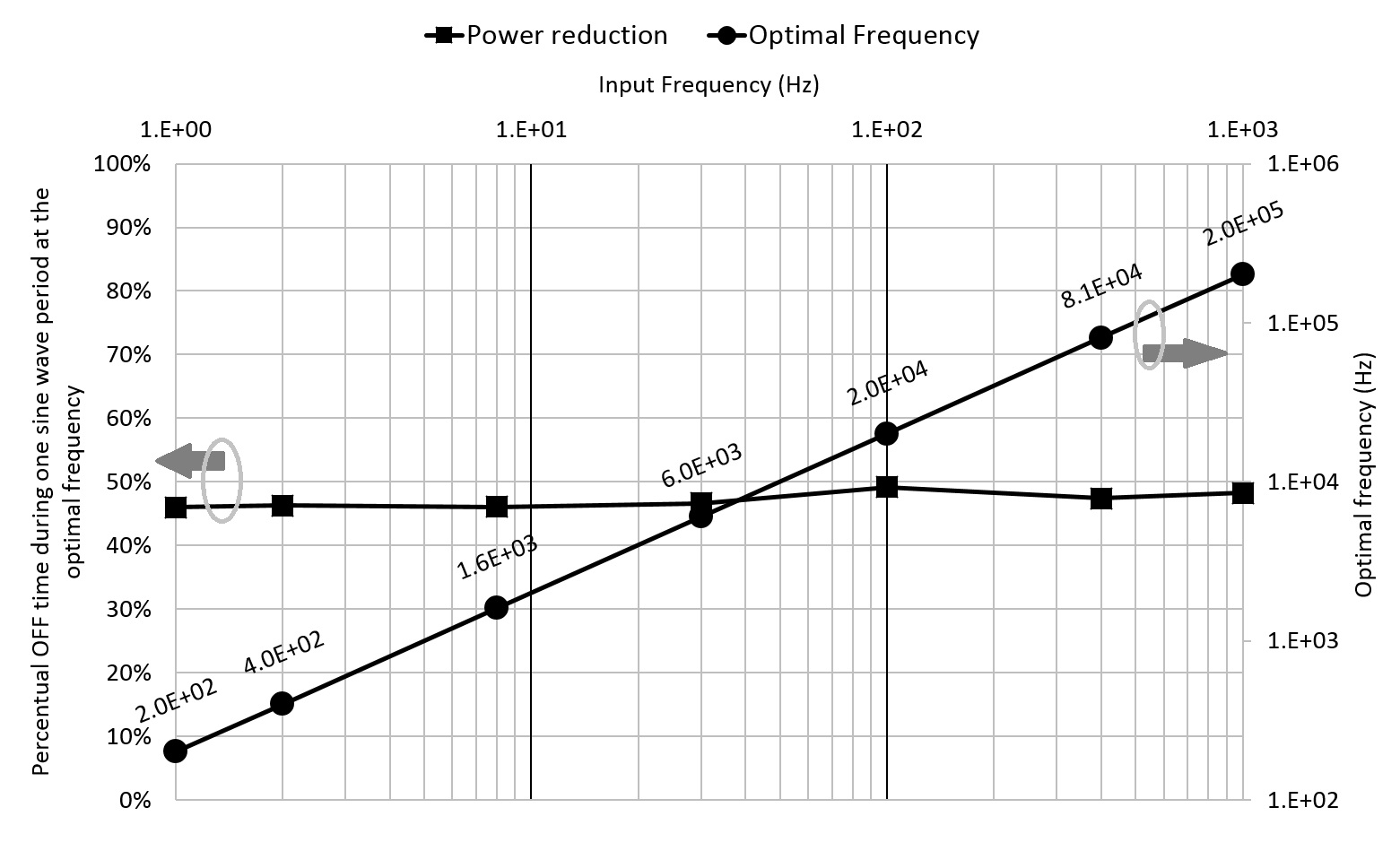}
    \caption{Optimal clock frequency for the maximum off time for the comparator at given input signal.}
    \label{curva_op_clk}
\end{figure}

The first intuition we can build from the expressions is the fact that the effectiveness of the technique for power reduction is better if the signal is more active, thus increasing the number of crossings per unit of time. To further guide the design choices for clock speed, some curves were build based on macro models of the converter and the control circuit. Figure \ref{curva_op_clk} shows the optimal clock choice for a full-scale sine wave driving the input of the converter at given frequency. Note that, for a full-scale signal, the converter cannot correctly handle frequencies higher that what is bounded by the clock value. At that maximum frequency, the percentage of off time of the comparators converge to a value close to $ 45.5\% $, where small variations are expected given the random distribution of the crossing moment and the time of the simulation. 

\begin{figure}[h!]
    \centering
    \includegraphics[width=85mm]{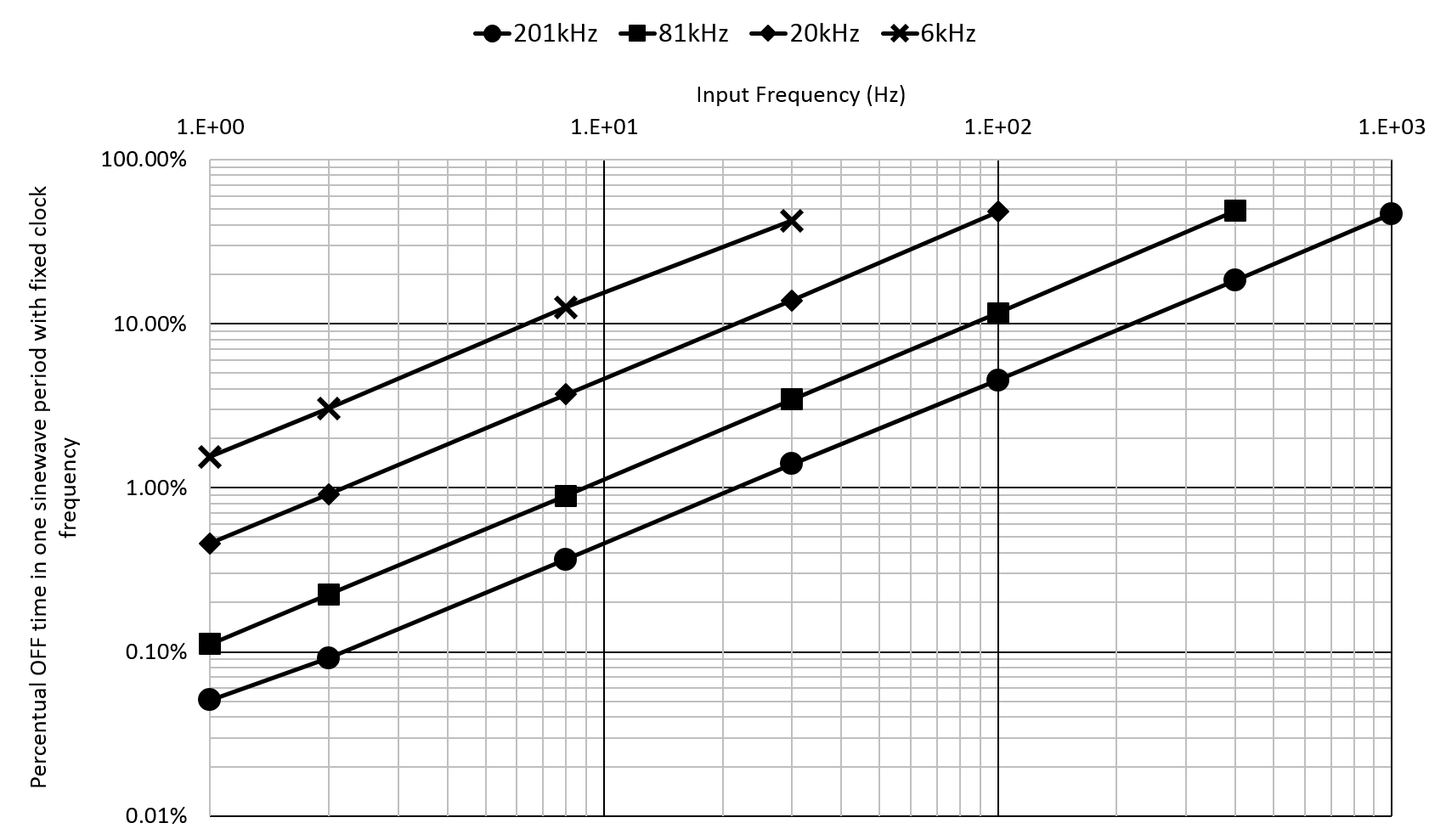}
    \caption{Decrease of off time for the comparators as input frequency decreases given reduction of the number of crossings at a fixed clock frequency.}
    \label{curva_reduc}
\end{figure}

If the clock frequency is sized and then fixed by design, then one can expect a decrease of the percentage of off time of the comparators, thus an increase in average power consumption given a smaller sine wave frequency at the input, as shown in Figure \ref{curva_reduc}. However, as small as the power reduction may be, the converter is always saving power for a clock frequency smaller than $201kHz$. This is due to the fact that the power overhead created by turning on and off the comparators and propagating the REQ signal in the simulations is smaller than the energy saved during the $5 \mu s$ off time of the given clock.

\begin{figure}[h!]
    \centering
    \includegraphics[width=85mm]{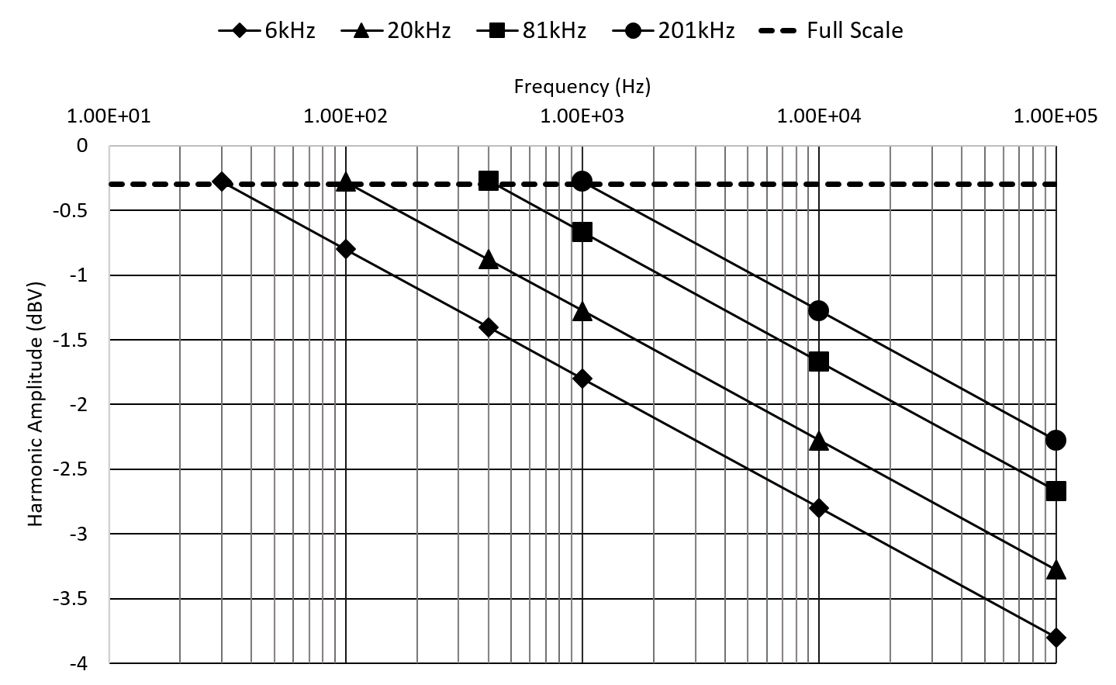}
    \caption{Bandwidth boundary for the proposed level crossing ADC given different clock frequencies.}
    \label{lv_bw}
\end{figure}

Finally, in Figure \ref{lv_bw}, one can see the bandwidth trade-off with respect to the chosen clock frequency. The top line is the limit set by the comparator input, the curves are then shaped by the Equation \ref{bandwidth}. Note the contrast with the brick-wall shaped bandwidth of Nyquist converters, here the bandwidth is hyperbolic shaped (trapezoidal in a logarithmic scale), thus bigger frequencies can be correctly converted if the amplitude of the input signal is correctly scaled. The power reduction gained from bandwidth trade-off is not straightforward since it is also dependent on the input signal. Table 1 presents a summary of the power performance of the proposed level crossing ADC under the power reduction technique.

\begin{table}[h]
\processtable{Summary of level crossing ADC simulation.}
{\begin{tabular}{|l|c|} \hline
     Parameter & Value  \\ \hline
     &\\[-0.8em]
     $P_{ADC}^{ON}$ & $2.6 \mu W$ \\ 
     &\\[-0.8em]\hline
     &\\[-0.8em]
     $P_{ADC}^{OFF}$ & $0.2 \mu W$ \\ 
     &\\[-0.8em]\hline
     &\\[-0.8em]
     $P_{ADC}^{average}$ at opt. clock frequency & $1.5 \mu W$ \\
     &\\[-0.8em]\hline
     Max. average power reduction (sine wave) & $42 \%$ \\ \hline
     Bandwidth (full scale input and 201kHz clock) & $1kHz$ \\ \hline
\end{tabular}}{}
\end{table}

\vfill\pagebreak

\section{Conclusion}
A modified architecture for level crossing ADC was presented and simulated. The comparator was designed to prevent glitches when powering up and a control circuit was built to define the time the comparators would be off after a level crossing. This enables the ADC to trade bandwidth with power savings by simply changing an input clock frequency. The free running ADC would consume an average power of $2.6 \mu W$ and with the new input dependent power consumption it could achieve an average power consumption $42\%$ smaller.

\vskip3pt
\ack{this research work was supported by Mini-ASIC and financed in part by the Coordena\c{c}\~ao de Aperfei\c{c}oamento de Pessoal de N\'ivel Superior - Brasil (CAPES) - Finance Code 001.}

\vskip5pt

\noindent L. M. Santana, D. L. de Oliveira and L. A. Faria (\textit{Instituto Tecnologico de Aeronautica, Sao Jose dos Campos, Brazil})
\vskip3pt

\noindent E-mail: lucasms@ieee.org

\end{document}